\documentclass{andp2012}% no class options needed by now
\usepackage[english]{babel}
\usepackage{epsfig}
\usepackage{graphicx}
\title{Quantum thermalization and vanishing thermal entanglement in the open Jaynes-Cummings model}
\author[L.-B. Fan]{Li-Bao Fan\inst{1}}
\author[Y.-H. Zhou]{Yue-Hui Zhou\inst{1}}
\author[ F. Zou]{ Fen Zou\inst{1}}
\author[H. Guo]{Huan Guo\inst{1}}
\author[J.-F. Huang]{Jin-Feng Huang\inst{1,}\footnote{Corresponding author\quad E-mail:~\textsf{jfhuang@hunnu.edu.cn}}}
\author[J.-Q. Liao]{ Jie-Qiao Liao\inst{1,}\footnote{Corresponding author\quad E-mail:~\textsf{jqliao@hunnu.edu.cn}}}
\address[1]{Key Laboratory of Low-Dimensional Quantum Structures and Quantum Control of
Ministry of Education, Key Laboratory for Matter Microstructure and Function of Hunan Province, Department of Physics and Synergetic Innovation Center for Quantum Effects and Applications, Hunan Normal University, Changsha 410081, China}
\shortauthors{F. Author et al.}
\begin{abstract}
The quantum thermalization of the Jaynes-Cummings (JC) model in both equilibrium and non-equilibrium open-system cases is sdudied, in which the two subsystems, a two-level system and a single-mode bosonic field, are in contact with either two individual heat baths or a common heat bath. It is found that in the individual heat-bath case, the JC model can only be thermalized when either the two heat baths have the same temperature or the coupling of the JC system to one of the two baths is turned off. In the common heat-bath case, the JC system can be thermalized irrespective of the bath temperature and the system-bath coupling strengths. The thermal entanglement in this system is also studied. A \emph{counterintuitive} phenomenon of \emph{vanishing} thermal entanglement in the JC system is found and proved.
\end{abstract}
\shortabstract
\begin{document}
\captionsetup[figure]{labelfont={bf},labelformat={default},labelsep=period,name={Figure}}
\captionsetup[Table]{labelfont={bf},labelformat={default},labelsep=period,name={Table}}
\maketitle
% \noindent

\section{Introduction}

The Jaynes-Cummings (JC) model,\textsuperscript{\cite{bib1}} which describes the interaction of a two-level system (TLS) with a single-mode bosonic field,\textsuperscript{\cite{bib2}} is considered as one of the most important and basic models in quantum optics.\textsuperscript{\cite{bib3}} The JC model not only plays a significant role in the understanding of matter-filed interactions in various branches of modern physics such as atomic physics, quantum optics, and solid-state physics, but also has wide applications in frontier quantum technologies including quantum control, quantum precision measurement, and quantum information processing.\textsuperscript{\cite{bib4,bib5}} So far, many important physical effects and phenomena have been observed in the JC model, such as the quantization of electromagnetic fields,\textsuperscript{\cite{bib6}} the vacuum Rabi splitting,\textsuperscript{\cite{bib7,bib8,bib9,bib10,bib11,bib12,bib13,bib14}} and the collapse and revival of the inversion population of the TLS.\textsuperscript{\cite{bib15,bib16}} All the phenomena observed are subjected to the influence of the dissipations because quantum systems are inevitably coupled to their environments. In particular, many interesting observations are based on the steady state of the JC system. Therefore, quantum statistical physics such as quantum thermalization\textsuperscript{\cite{bib17,bib18}} and thermal entanglement\textsuperscript{\cite{bib19,bib20}} are significant research topics in the open JC model.

With the development of experimental techniques, the JC model can be implemented with more and more physical systems.\textsuperscript{\cite{bib3}} As a result, the couplings of the JC system with its environments are enriched because the TLS and the bosonic mode of the JC model could be in contact with either two individual heat baths (IHBs) or a common heat bath (CHB). In most previous studies, the dissipations of the JC model are introduced by phenomenologically adding the independent Lindblad dissipators for the TLS and the bosonic mode into the Liouville equation.\textsuperscript{\cite{bib17,bib21,bib22}} This treatment will lead to unphysical results when the coupling strength between the subsystems is much larger than their decay rates.\textsuperscript{\cite{bib23}} Therefore, a microscopic derivation of quantum master equation\textsuperscript{\cite{bib23,bib24,bib25,bib26,bib27,bib28,bib29}} describing the evolution of the JC model in both the IHB and CHB cases becomes an important and desired task.\textsuperscript{\cite{bib30,bib31,bib32}} We note that some attention has been paid to the study of equilibrium and non-equilibrium steady states in composite quantum systems by deriving quantum master equations in the dressed-state representation of the coupled systems.\textsuperscript{\cite{bib33,bib34,bib35,bib36,bib37,bib38,bib39}} For example, quantum thermalization and entanglement of two coupled two-level atoms have been studied in the dressed-state representation,\textsuperscript{\cite{bib33}} and the quantum statistical properties of the open quantum Rabi model have been studied in the eigenstate representation.\textsuperscript{\cite{bib37}} In addition, many topics in statistical physics have been studied based on the microscopically derived quantum master equation in coupled-atom
systems\textsuperscript{\cite{bib40,bib41,bib42,bib43,bib44,bib45,bib46,bib47,bib48,bib49}} and coupled-harmonic-oscillator systems.\textsuperscript{\cite{bib30,bib50}}

In this paper, we study quantum thermalization and thermal entanglement of the JC model in either the IHB or the CHB case. We treat the JC model as an effective multi-level system and derive the quantum master equations in the eigenstate representation. We introduce the effective temperatures associated with any two eigenstates, and study the thermalization of the JC model by inspecting whether the steady-state density matrix of the JC system can be written as a thermal equilibrium state. In the IHB case, we find that when the two IHBs have the same (different) temperatures, the JC model can (cannot) be thermalized with the same temperature as those of these two IHBs. In particular, when the coupling of one of two subsystems with the corresponding heat bath is turned off, the JC system can be thermalized with the contacted heat bath. In the CHB case, the JC system can reach a thermal equilibrium with the common bath. In addition, we study quantum entanglement between the TLS and the bosonic mode by calculating the logarithmic negativity of the thermal equilibrium state. We find and show a \emph{counterintuitive} phenomenon of \emph{vanishing} thermal entanglement in the JC system.

%%%%%%%%%%%%%%%%%%%%%
\begin{figure}
\center
\includegraphics[bb=0 0 280 329, width=0.47 \textwidth]{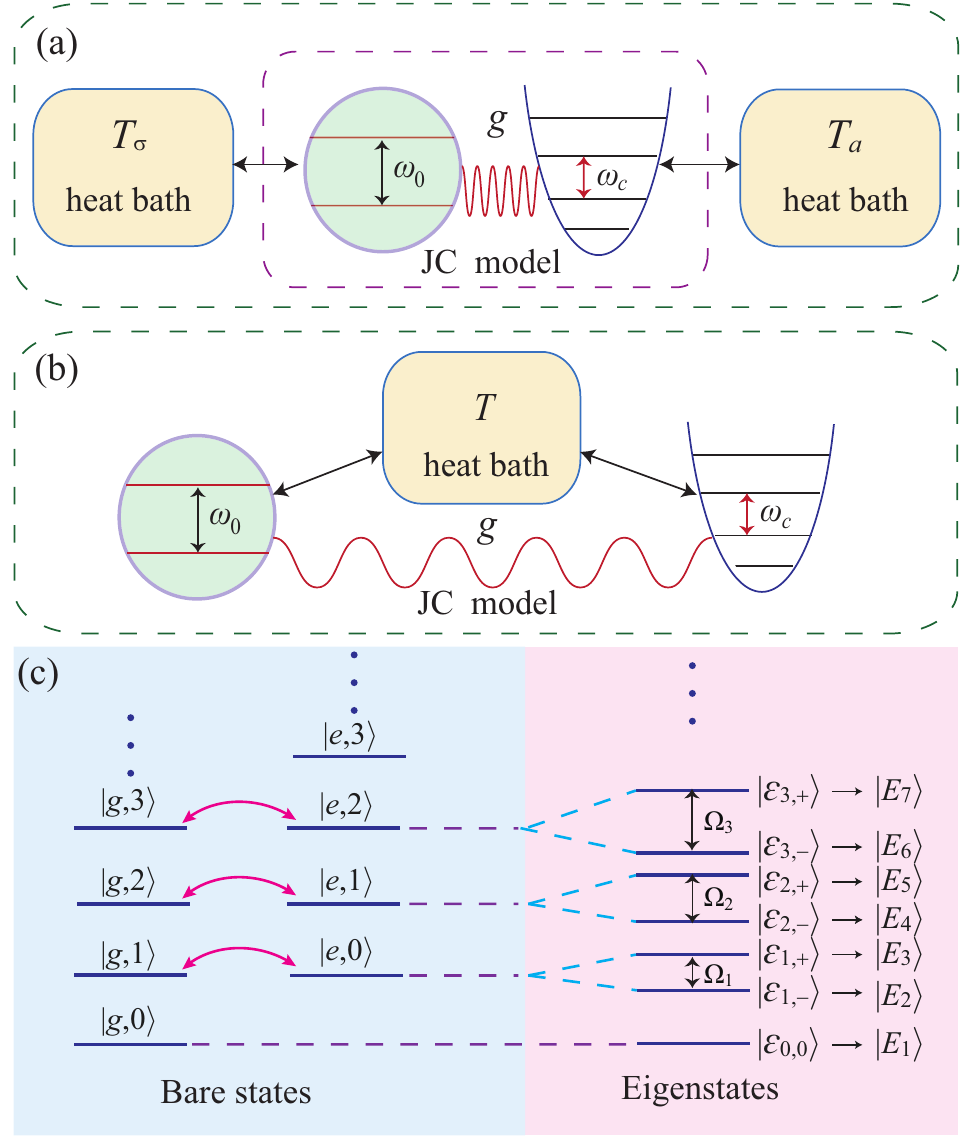}
\caption{Schematic diagram of the open JC model in the a) individual and b) common heat-bath cases, in which the two subsystems: a two-level system and a single-mode bosonic field, are coupled to two individual heat baths or a common heat bath. c) The bare states and eigenstates (dressed states) of the JC model.}\label{Fig1}
\end{figure}
%%%%%%%%%%%%%%%%%%%%%

\section{Model and Hamiltonian \label{Sec:Model}}

We consider the JC model (Figure~\ref{Fig1}), which describes the interaction of a TLS with a single-mode bosonic field. The Hamiltonian of the JC model reads\textsuperscript{\cite{bib1}}
\begin{equation}
H_\mathrm{JC}=\frac{\hbar\omega_{0}}{2}\sigma_{z}+\hbar\omega_{c} a^{\dagger}a+\hbar g(a^{\dagger}\sigma_{-}+\sigma_{+}a).  \label{HJC}
\end{equation}
Here the TLS is described by the Pauli operators $\sigma_{x}$=$\vert e\rangle\langle g\vert+\vert g\rangle\langle e\vert $, $\sigma _{y}=i(\vert g\rangle\langle e\vert-\vert e\rangle\langle g\vert) $, and $\sigma_{z}$=$\vert e\rangle\langle e\vert-$\\$\vert g\rangle\langle g\vert$, which are defined based on the excited state $\vert e\rangle $ and the ground state $\vert g\rangle$, with the energy separation $\hbar\omega_{0}$. The raising and lowering operators in equation~(\ref{HJC}) are defined by $\sigma_{\pm}$=$(\sigma_{x}\pm i\sigma_{y})/2$. The operator $a$ ($a$$^{\dagger}$) is the annihilation (creation) operator of the bosonic field with resonance frequency $\omega_{c}$. The last term in equation~(\ref{HJC}) describes the JC-type interaction between the TLS and the bosonic mode, with $g$ being the coupling strength.

In a system depicted by the JC model, the total excitation number operator $N=a^{\dagger}a+\sigma_{+}\sigma_{-}$ is a conserved quantity based on the commutative relation $[N,H_\mathrm{JC}]=0$. Therefore, the whole Hilbert space of the system can be decomposed into a series of subspaces with different excitation numbers $n$ ($n=0,1,2,3,\cdots$). In the $n$-excitation subspace, the eigenequation of the Hamiltonian can be expressed as
$H_\mathrm{JC}\vert\varepsilon_{n,\alpha_{n}}\rangle=\hbar\varepsilon_{n,\alpha_{n}}\vert\varepsilon_{n,\alpha_{n}}\rangle,\hspace{0.3cm} n=0,1,2,\cdots,$ where $\alpha_{0}=0$ for $n=0$ and $\alpha_{n}=\pm$ for $n\geq1$. In the zero-excitation subspace, the eigenstate is $\vert\varepsilon_{0,0}\rangle=\vert g, 0\rangle$\ and the eigenenergy is $\varepsilon _{0,0}=-{\hbar\omega_{0}}/{2}$. In the nonzero $n$-excitation subspace, the eigenstates and eigenvalues are defined by $\vert \varepsilon_{n,+}\rangle=\cos(\theta_{n}/2)\vert e\rangle\vert n-1\rangle+\sin(\theta_{n}/2)\vert g\rangle\vert n\rangle$, $\vert \varepsilon_{n,-}\rangle=-\sin(\theta_{n}/2)\vert e\rangle\vert n-1\rangle+\cos(\theta_{n}/2)\vert g\rangle\vert n\rangle$, and $\varepsilon_{n,\pm}=\hbar\omega_{c}(n-1/2)\pm\hbar\Omega_{n}/2$, where the mixing angle $\theta_{n}$ is defined by $\tan\theta_{n} =2g\sqrt{n}/\delta$, and the Rabi frequency $\Omega_{n}=\sqrt{\delta^{2}+4 g^{2}n}$ is introduced, with $\delta=\omega_{0}-\omega_{c}$  being the detuning.

In terms of these eigenstates, the JC Hamiltonian can be expressed in the eigenstate representation as $H_\mathrm{JC}=\sum_{n=0}^{\infty}\sum_{\alpha_{n}}\hbar\protect\varepsilon_{n,\alpha_{n}}\vert\protect\varepsilon_{n,\alpha_{n}}\rangle\langle\protect\varepsilon_{n,\alpha_{n}}\vert$.
We note that many interesting physical effects in the JC system can be explained based on its eigensystem. For example, (i) the transitions from states  $\vert\varepsilon_{1,\pm}\rangle$ to $\vert\varepsilon_{0,0}\rangle$ correspond to two peaks in the vacuum Rabi splitting spectrum. (ii) The JC model predicts a characteristic nonlinearity of $\sqrt{n}$ in the resonant case. These effects have been demonstrated in cavity-QED \textsuperscript{\cite{bib5}} and circuit-QED\textsuperscript{\cite{bib10,bib14,bib51}} systems.

In a realistic situation, the TLS and the bosonic field couple inevitably with the environments surrounding them. In this paper, we shall consider two kinds of different cases: one is the IHB case [Figure~\ref{Fig1}a] and the other is the CHB case [Figure~\ref{Fig1}b]. In the IHB case, the TLS and the bosonic mode are coupled to two individual heat baths. The free Hamiltonian of the two baths reads
$H^{(\textrm{IHB})}_{B}=\sum_{q}\hbar\omega_{q}c_{q}^{\dagger}c_{q}+\sum_{k}\hbar\omega_{k}b_{k}^{\dagger}b_{k}$,
where the creation and annihilation operators $c^{\dag}_{q}$ ($b^{\dag}_{k}$) and $c_{q}$ ($b_{k}$) describe the $q$th ($k$th) harmonic oscillator with frequency $\omega_{q}$ ($\omega_{k}$) in the heat bath of the TLS (bosonic mode). The interaction Hamiltonian of the JC model with the two baths reads
\begin{equation}
H^{(\textrm{IHB})}_{I}=\sum_{q}\hbar\lambda_{q}\sigma_{x}(c_{q}^{\dagger}+c_{q})+\sum_{k}\hbar\eta_{k}(a^{\dagger}+a)(b_{k}^{\dagger}+b_{k}),\label{HintIHB}
\end{equation}
where $\lambda_{q}$ ($\eta_{k}$) is the coupling strength between the TLS (bosonic field) and the $q$th ($k$th) mode of its bath.

In the CHB case, as shown in Figure~\ref{Fig1}b, the TLS and the bosonic mode are immersed in a CHB with the Hamiltonian
$H^{(\textrm{CHB})}_{B}=\sum_{k}\hbar\omega_{k}c_{k}^{\dagger}c_{k}$,
where  $c^{\dag}_{k}$  and $c_{k}$ are, respectively, the creation and annihilation operators of the $k$th harmonic oscillator with the resonance frequency $\omega_{k}$ of the CHB. The interaction Hamiltonian between the JC system and the CHB reads
\begin{equation}
H^{(\textrm{CHB})}_{I}=\sum_{k}\hbar\lambda_{k}\sigma_{x}(c_{k}^{\dagger}+c_{k})+\sum_{k}\hbar\eta_{k}(a^{\dagger}+a)(c_{k}^{\dagger}+c_{k}),\label{nondiacouplingH}
\end{equation}
where $\lambda_{k}$ ($\eta_{k}$) is the coupling strength between the TLS (bosonic field) and the $k$th mode of the CHB.

In both the IHB and CHB cases, the total Hamiltonian of the system can be written as
$H=H_\mathrm{JC}+H^{(s)}_{B}+H^{(s)}_{I}$ for $s=\text{``IHB"}$ and $\text{``CHB"}$,
which is schematically shown in Figures~\ref{Fig1}a and~\ref{Fig1}b, where $H_\mathrm{JC}$ is the JC Hamiltonian, $H^{(s)}_{B}$ and $H^{(s)}_{I}$ are the bath Hamiltonians and the interaction Hamiltonians between the JC system and its baths in the IHB and CHB cases, respectively.

\section{Quantum thermalization in the IHB case \label{Sec:IHB}}

In this section, we study quantum thermalization of the JC model in the IHB case. A dressed-state quantum master equation will be derived to govern the evolution of the JC model. By introducing effective temperatures, we will analyze the thermalization problem of the JC system.

\subsection{Quantum master equation}

To describe the damping effects in this system, we will derive the quantum master equation by employing the standard Born-Markov approximation under the condition of weak system-bath couplings and short bath correlation times.\textsuperscript{\cite{bib17}} In particular, we will derive the quantum master equation in the eigenstate representation of the JC Hamiltonian. When the TLS and the bosonic mode are coupled to two individual heat baths, the coupling between the JC system with their baths is described by Hamiltonian~(\ref{HintIHB}). Below, we will derive the quantum master equation in the interaction picture with respect to the free Hamiltonian $H_{0} =H_{\text{JC}}+H^{\text{IHB}}_{B}$. Within the Born-Markov framework, the equation for the reduced density matrix of the system in the interaction picture can be written as\textsuperscript{\cite{bib17}}
\begin{equation}
\frac{d}{dt}\tilde{\rho}_{S}(t)=-\int_{0}^{\infty }d{t}^{\prime}\mathrm{Tr}_{B}[H_{I}(t),[H_{I}(t-{t}^{\prime}),\tilde{\rho}_{S}(t)\otimes\rho_{B}]],\label{master-equt}
\end{equation}
where $\tilde{\rho}_{S}(t)$ is the reduced density matrix of the JC system in the interaction picture. The system-bath interaction Hamiltonian in the interaction picture is defined by $H_{I}(t)=e^{iH_{0}t/\hbar}H^{\text{IHB}}_{I}e^{-iH_{0}t/\hbar}$. By substituting the Hamiltonians $H_{I}(t)$ and $H_{I}(t-{t}^{\prime})$ into equation~(\ref{master-equt}) and making the secular approximation, we can obtain the quantum master equation in the interaction picture as

\begin{eqnarray}
\frac{d}{dt}\tilde{\rho}_{S}(t)=L_{\text{IHB}}[\tilde{\rho}_{S}(t)]
&=&\sum_{n=0}^{\infty}\sum_{\alpha_{n},\beta_{n}}\sum_{l=\sigma,a}\frac{1}{2}\gamma_{l}(\omega_{n,\alpha_{n+1},\beta_{n}})\nonumber\\
&&\times\vert\chi_{l,\alpha_{n+1},\beta_{n}}\vert^{2}[\bar{n}_{l}(\omega_{n,\alpha_{n+1},\beta_{n}})+1]\nonumber\\
&&\times\mathcal{D}[\vert\varepsilon_{n,\beta_{n}}\rangle\langle\varepsilon_{n+1,\alpha_{n+1}}\vert]\tilde{\rho}_{S}(t) \nonumber\\
&&+\sum_{n=0}^{\infty}\sum_{\alpha_{n},\beta_{n}}\sum_{l=\sigma,a}\frac{1}{2}\gamma_{l}(\omega_{n,\alpha_{n+1},\beta_{n}})\nonumber\\
&&\times\vert\chi_{l,\alpha_{n+1},\beta_{n}}\vert^{2}\bar{n}_{l}(\omega_{n,\alpha_{n+1},\beta_{n}})\nonumber\\
&&\times\mathcal{D}[\vert\varepsilon_{n+1,\alpha_{n+1}}\rangle\langle\varepsilon_{n,\beta_{n}}\vert]\tilde{\rho}_{S}(t),    \label{LindbladIHB}
\end{eqnarray}
with the Lindblad superoperator $\mathcal{D}[{O}]\tilde{\rho}_{S}(t)$ defined by\\$\mathcal{D}[{O}]\tilde{\rho}_{S}(t)=2{O}\tilde{\rho}(t)_{S}{O}^{\dagger}-\tilde{\rho}_{S}(t){O}^{\dagger }{O}-{O}^{\dagger}{O}\tilde{\rho}_{S}(t)$.

In equation~(\ref{LindbladIHB}), the decay rates related to the dissipation channels of the TLS and the bosonic mode are defined by
\begin{subequations}
\begin{align}
\gamma_{\sigma}(\omega_{n,\alpha_{n+1},\beta_{n}})=&2\pi\varrho_{\sigma}(\omega_{n,\alpha_{n+1},\beta_{n}})\lambda^{2}(\omega_{n,\alpha_{n+1},\beta_{n}}), \\ \gamma_{a}(\omega_{n,\alpha_{n+1},\beta_{n}})=&2\pi\varrho_{a}(\omega_{n,\alpha_{n+1},\beta_{n}})\eta^{2}(\omega_{n,\alpha_{n+1},\beta_{n}}),
\end{align}
\end{subequations}
where $\varrho_{\sigma}(\omega_{q})$ and $\varrho_{a}(\omega_{k})$ are, respectively, the spectral density functions of the heat baths in contact with the TLS and the bosonic mode, and $\omega_{n,\alpha_{n+1},\beta_{n}}=\varepsilon_{n+1,\alpha_{n+1}}-\varepsilon_{n,\beta_{n}}$ represent the energy separation between the two eigenstates $\vert\varepsilon_{n+1,\alpha_{n+1}}\rangle$ and $\vert\varepsilon_{n,\beta_{n}}\rangle$ of the JC Hamiltonian. In our simulations, we assume that the decay rates $\gamma_{\sigma}(\omega_{n,\alpha_{n+1},\beta_{n}})=\gamma_{\sigma}$ and $\gamma_{a}(\omega_{n,\alpha_{n+1},\beta_{n}})=\gamma_{a}$, which means that the decay rates associated with all the transitions induced by the same subsystem are identical.

The transition coefficients in equation~(\ref{LindbladIHB}) induced by the system-environment coupling terms can be calculated as $\chi_{\sigma,\alpha_{n+1},\beta_{n}}=\langle\varepsilon_{n+1,\alpha_{n+1}}|\sigma_{x}\vert\varepsilon_{n,\beta_{n}}\rangle$, $\chi_{a,\alpha_{n+1},\beta_{n}}=\langle\varepsilon_{n+1,\alpha_{n+1}}|(a+a^\dagger)\vert\varepsilon_{n,\beta_{n}}\rangle$. The average thermal excitation numbers in equation~(\ref{LindbladIHB}) are defined by $\bar{n}_{s=\sigma,a}(\omega)=1/[\exp(\hbar\omega/k_{B}T_{s})-1]$, where $k_{B}$ is the Boltzmann constant, $\hbar\omega$ denotes the energy separation associated with the two eigenstates involved, and $T_{\sigma}$ and $T_{a}$ are the temperatures of the heat baths of the TLS and the bosonic mode, respectively.

According to the transformation $\rho_{S}(t)=e^{-iH_{\text{JC}}t/\hbar}\tilde{\rho}_{S}(t)\\e^{iH_{\text{JC}}t/\hbar}$ and quantum master equation in the interaction picture, the quantum master equation in the Schr\"{o}dinger picture can be obtained as
\begin{equation}
\frac{d}{dt}\rho_{S}(t)=-\frac{i}{\hbar}[H_\mathrm{JC},\rho_{S}(t)]+L_{\text{IHB}}[\rho_{S}(t)],   \label{MasteqIHBSP}
\end{equation}
where $\rho_{S}(t)$ is the reduced density matrix of the JC system in the Schr\"{o}dinger picture, and the dissipator $L_{\text{IHB}}[\rho_{S}(t)]$ has the same form as equation~(\ref{LindbladIHB}) under the replacement of $\tilde{\rho}_{S}(t)\rightarrow\rho_{S}(t)$.

In this system, the system-environment couplings will induce transitions between the eigenstates of the JC model. The transition selection rule between these eigenstates can be found by calculating the transition matrix elements of the operators $\sigma_{x}$ and $(a+a^{\dagger})$ in the eigenstate representation. As an example, below we analyze the transition matrix elements in the resonant case $\delta=0$.  We obtain these matrix elements between the zero-excitation state and the two one-excitation eigenstates as
$\langle\varepsilon_{1\pm}\vert\sigma_{x}\vert\varepsilon_{0,0}\rangle =\pm1/\sqrt{2}$ and $
\langle\varepsilon_{1\pm}\vert(a^{\dag}+a)\vert\varepsilon_{0,0}\rangle=1/\sqrt{2}$.
Similarly, the transition matrix elements between the eigenstates in the $n$- and $(n+1)$-excitation ($n>0$) subspaces can be obtained as
\begin{subequations}
\begin{align}
\langle\varepsilon_{m+}\vert\sigma_{x}\vert\varepsilon_{n\pm}\rangle=&\frac{1}{2}(\delta_{n,m-1}\pm\delta_{n,m+1}),\\
\langle\varepsilon_{m-}\vert\sigma_{x}\vert\varepsilon_{n\pm}\rangle=&\frac{1}{2}(-\delta_{n,m-1}\pm\delta_{n,m+1}),\\
\langle\varepsilon_{m+}\vert(a^{\dag}+a)
\vert\varepsilon_{n\pm}\rangle=&\frac{1}{2}[(\sqrt{m}\pm\sqrt{m-1})\delta_{n,m-1}\nonumber\\
&+(\sqrt{m+1}\pm\sqrt{m})\delta_{n,m+1}],\\
\langle\varepsilon_{m-}\vert(a^{\dag }+a)
\vert\varepsilon_{n\pm}\rangle=&\frac{1}{2}[(\sqrt{m}\mp\sqrt{m-1})\delta_{n,m-1}\nonumber\\
&+(\sqrt{m+1}\mp\sqrt{m})\delta_{n,m+1}].
\end{align}
\end{subequations}

Here, we can see that the transitions induced by the system-environment couplings can only take place between the eigenstates in the neighboring excitation-number subspaces, this is because the operators $\sigma_{x}$ and $(a+a^{\dagger})$ in the system-environment coupling terms lead to the change of one excitation in the transition processes. In addition, the two states in the $n$-excitation subspace will be coupled to all the states in both the $(n-1)$- and $(n+1)$-excitation subspaces.

\subsection{Quantum thermalization \label{ThermIHB}}
%%%%%%%%%%%%%%%%%%%%%
\begin{figure}
\center
\includegraphics[width=0.47 \textwidth]{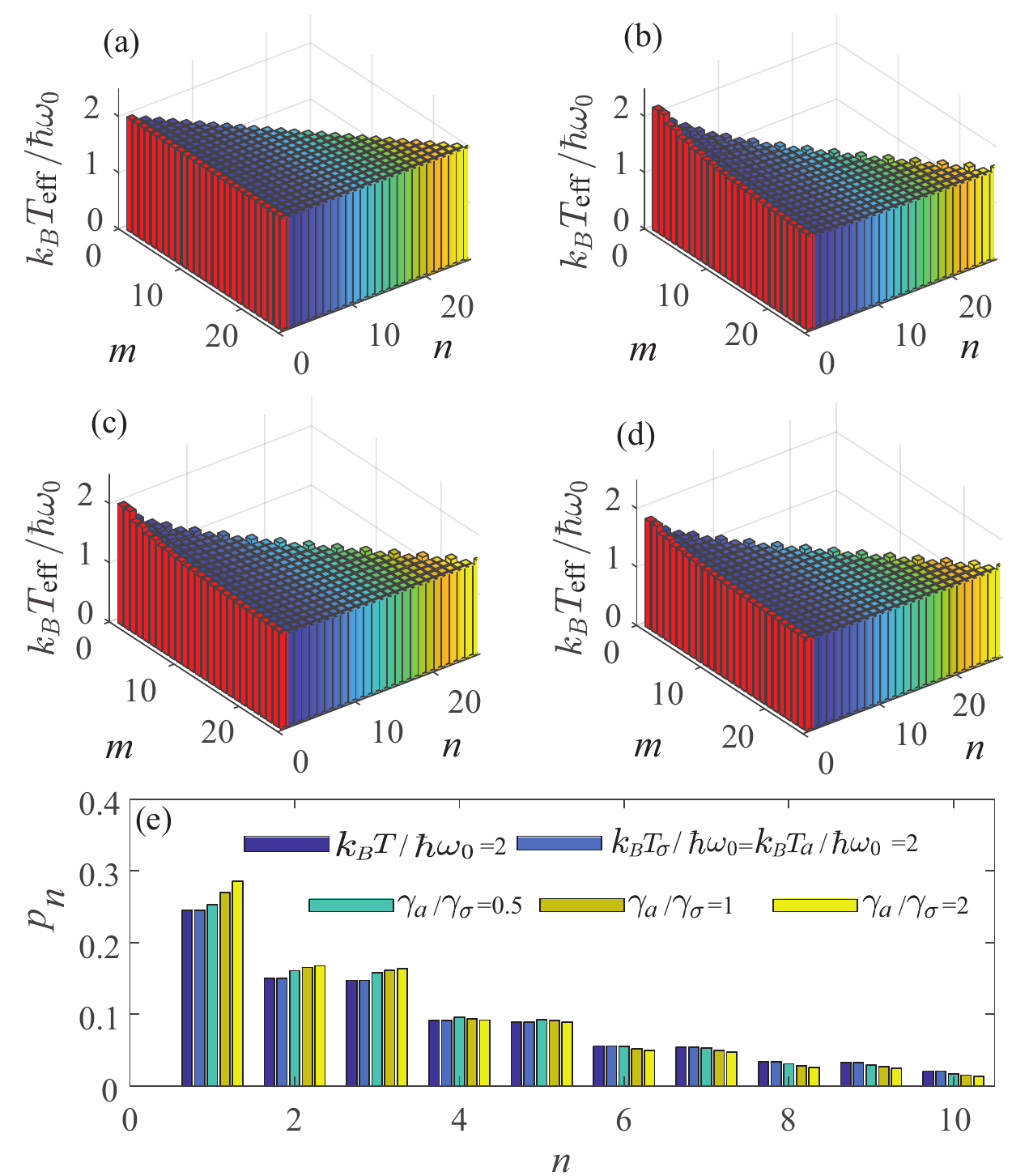}
\caption{\label{Fig2}The scaled effective temperatures $k_{B}T_\text{eff}(\omega_{m,n})\\/\hbar\omega_{0}$  as functions of the energy-level indexes $m$ and $n$ in various cases: a) $k_{B}T_{a}/\hbar\omega_{0}=k_{B}T_{\sigma}/\hbar\omega_{0}=2$ and $\gamma_{\sigma}/\omega_{0}=\gamma_{a}/\omega_{0}=0.0001$. b)-d) $k_{B}T_{a}/\hbar\omega_{0}=1.5$, $k_{B}T_{\sigma}/\hbar\omega_{0}=2.5$, $\gamma_{\sigma}/\omega_{0}=0.0001$, and $\gamma_{a}/\gamma_{\sigma}=0.5$, $1$, and $2$. e) The steady-state populations $p_{n}$ of the eigenstates $\vert E_{n}\rangle$.  The parameters of these bars correspond to these four cases in panels (a)-(d), respectively. The blue (first group) bars represent the populations of a thermal state at $k_{B}T/\hbar\omega_{0}=2$. The five contiguous color bars represent the distribution of the same eigenstate $|E_{n}\rangle$ in five different situations. Other used parameters are $\omega_{a}/\omega_{0}=1$ and $g/\omega_{0}=0.02$. }
\end{figure}
%%%%%%%%%%%%%%%%%%%%%

We now turn to the study of quantum thermalization of the JC system based on the steady-state solution of quantum master equation~(\ref{MasteqIHBSP}). It is well known that the quantum thermalization\textsuperscript{\cite{bib17}} is an irreversibly dynamic process via which a quantum system approaches a thermal equilibrium with the same temperature as that of the bath surrounding it. The JC system in a thermal equilibrium at temperature $T$ is described by the density matrix of the thermal state $\rho_{\text{th}}(T)=Z_\text{JC}^{-1}\exp[-H_\text{JC}/(k_{B}T)]$, which depends on the Hamiltonian $H_\text{JC}$ and the bath temperature $T$, where $Z_\text{JC}=\textrm{Tr}_\text{JC}\{\exp[-H_\text{JC}/(k_{B}T)]\}$ is the partition function of the thermalized JC system. During the process of a quantum thermalization, all the initial-state information of the thermalized system is totally erased by its environment. In the present system, when the coupling strength between the TLS and the bosonic mode is stronger than the system-bath couplings, the physical system should be regarded as an effective multiple-level system (i.e., the JC system in the eigenstate representation) coupled to two IHBs. The dynamical evolution of the JC system approaching to its steady state can be understood as a non-equilibrium quantum thermalization: thermalization of a multi-level quantum system coupled to two IHBs at the the same or different temperatures.\textsuperscript{\cite{bib52,bib53,bib54,bib55}} In the steady state, the JC model is in a completely mixed state in the eigenstate representation. So we can introduce some effective temperatures to characterize the relation between any two eigenstates according to their steady-state populations. If all these temperatures defined based on these energy levels are the same, then we regard as the thermalization of the JC system. In this case, the density matrix of the system can be described by the thermal state $\rho_{\text{th}}(T)$.

To facilitate the marking of the eigenstates, we mark these eigenstates of the JC model as $\vert E_{n}\rangle$ starting from the lowest energy level, namely $\vert\varepsilon_{0,0}\rangle\rightarrow\vert E_{1}\rangle$, $\vert\varepsilon_{1,-}\rangle\rightarrow\vert E_{2}\rangle$, $\vert\varepsilon_{1,+}\rangle\rightarrow\vert E_{3}\rangle$, $\cdots$, $\vert\varepsilon_{n,-}\rangle\rightarrow\vert E_{2n}\rangle$, $\vert\varepsilon_{n,+}\rangle\rightarrow\vert E_{2n+1}\rangle$, $\cdots$.
We also denote the energy separation between the two eigenstates $\vert E_{m}\rangle$ and $\vert E_{n}\rangle$ as $\hbar\omega_{mn}=E_{m}-E_{n}$, and the populations of the states $\vert E_{m}\rangle$ and $\vert E_{n}\rangle$ as $p_{m}$ and $p_{n}$. Then, it is natural to define the frequency-dependent effective temperature as
$T_\mathrm{eff}(\omega_{mn})=(\hbar\omega_{mn}/k_{B})[\ln(p_{m}/p_{n})]^{-1}$.
%%%%%%%%%%%%%%%%%%%%%
\begin{figure}
\center
\includegraphics[width=0.47 \textwidth]{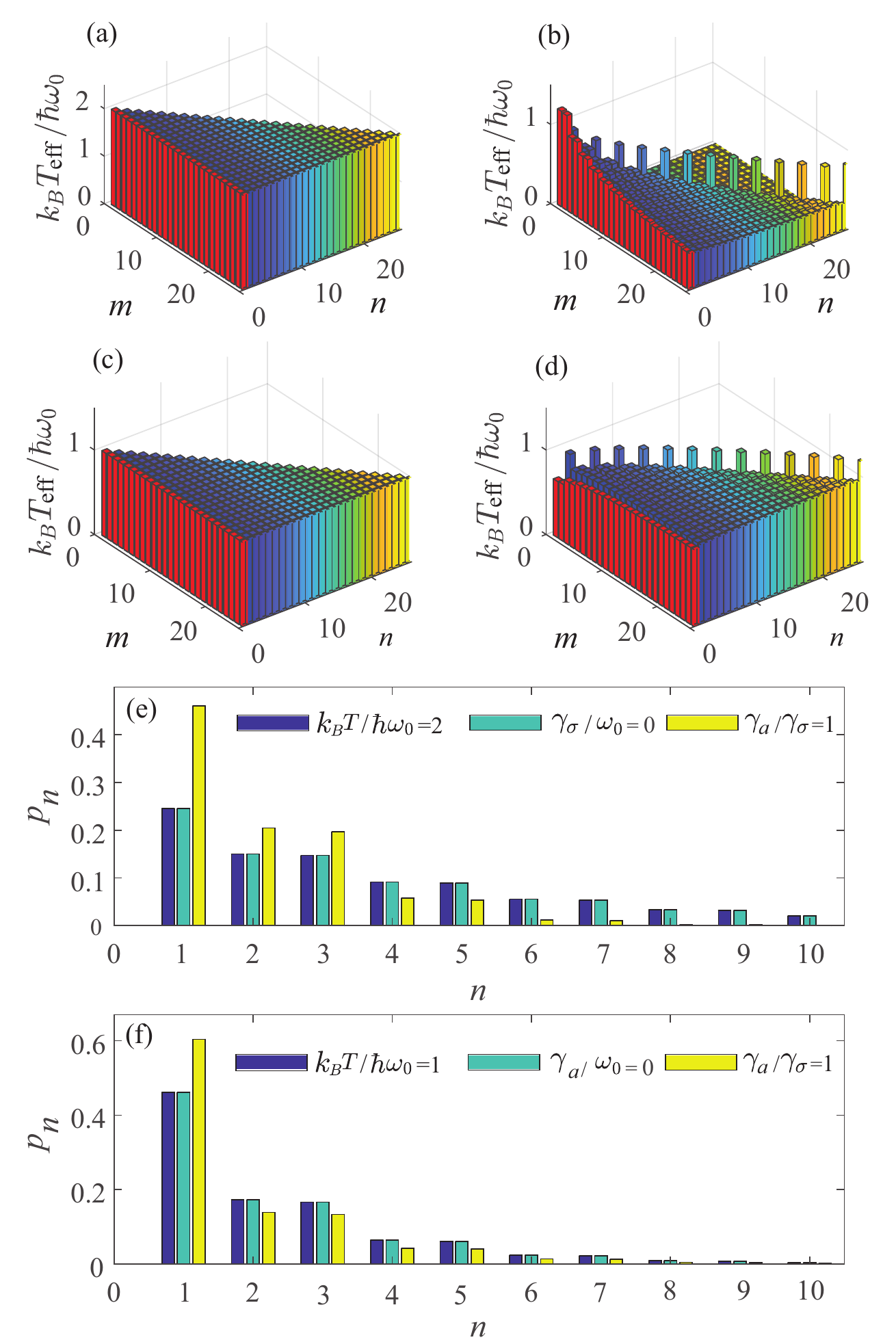}
\caption{\label{Fig3} The scaled effective temperatures $k_{B}T_\text{eff}(\omega_{m,n})\\/\hbar\omega_{0}$ as functions of the energy-level indexes $m$ and $n$ when  a) $\gamma_{a}/\omega_{0}=0$, $\gamma_{\sigma}/\omega_{0}=0.0001$, $k_{B}T_{\sigma}/\hbar\omega_{0}=2$, b) $\gamma_{a}/\omega_{0}=\gamma_{\sigma}/\omega_{0}=0.0001$, $k_{B}T_{a}/\hbar\omega_{0}=0$, $k_{B}T_{\sigma}/\hbar\omega_{0}=2$, c) $\gamma_{\sigma}/\omega_{0}=0$, $\gamma_{a}/\omega_{0}=0.0001$, $k_{B}T_{a}/\hbar\omega_{0}=1$, and d) $\gamma_{a}/\omega_{0}=\gamma_{\sigma}/\omega_{0}=0.0001$, $k_{B}T_{a}/\hbar\omega_{0}=1$, $k_{B}T_{\sigma}/\hbar\omega_{0}=0$. e) The steady-state populations $p_{n}$ of the eigenstates $\vert E_{n}\rangle$, the parameters of green and yellow bars correspond to these cases of (a) and (b), respectively. The blue bars (the first group) correspond to the populations of the thermal state at $k_{B}T/\hbar\omega_{0}=2$. f) The steady-state populations $p_{n}$ of the eigenstates $\vert E_{n}\rangle$, the parameters of green and yellow bars correspond to these cases of (c) and (d), respectively. The blue bars (the first group) correspond to the populations of the thermal state at $k_{B}T/\hbar\omega_{0}=1$. Other used parameters are the same as those given in Figure~\ref{Fig2}.}
\end{figure}
%%%%%%%%%%%%%%%%%%%%%%

Based on these effective temperatures, we can characterize the thermalization of the JC system. By numerically solving the equations of motion of these density matrix elements $\langle E_{n}|\rho_\text{ss}|E_{m}\rangle$ based on equation~(\ref{MasteqIHBSP}) under the replacement of $\frac{d}{dt}\rho_{s}(t)\rightarrow 0$ with Mathematica, we get the steady-state density operator elements $\langle E_{n}|\rho_\text{ss}|E_{m}\rangle$ of the system, then the population $p_{n}=\langle E_{n}|\rho_\text{ss}|E_{n}\rangle$ of the eigenstate $\vert E_{n}\rangle$ and the effective temperatures $T_\mathrm{eff}(\omega_{mn})$ can be calculated accordingly. In our realistic numerical simulations, we need to truncate the dimension of the Hilbert space of the bosonic field so that $\sum_{n=1}^{n_{d}}\langle E_{n}|\rho_\text{ss}|E_{n}\rangle$ is approximately normalized. Here we choose the truncation dimension of the bosonic field as $n_{d}=17$.

In Figure~\ref{Fig2}, we plot the effective temperatures $T_\mathrm{eff}(\omega_{mn})$ as functions of the energy-level indexes $m$ and $n$ when the bath temperatures $T_{\sigma}$ and $T_{a}$ take various values. Figures~\ref{Fig2}a and~\ref{Fig2}b-d correspond to the cases of $T_{\sigma}=T_{a}$ and $T_{\sigma}\neq T_{a}$, respectively. We find that the JC system can be thermalized when the two heat baths have the same temperatures $T_{\sigma}=T_{a}$, as shown in Figure~\ref{Fig2}a. In this case, the temperature of the thermalized JC system is the same as those of the two baths, and the density matrix of the JC system can be written as the thermal state. In particular, the thermalization of the system is independent of the nonzero values of the decay rates. When $T_{\sigma}\neq T_{a}$ (see Figures~\ref{Fig2}b-d), the system cannot be thermalized because these temperatures associated with these energy-level pairs have different values, and hence the steady state cannot be expressed as the thermal equilibrium density matrix. In order to understand the thermalization in the JC system, we plot the steady-state populations $p_{n}$ of the JC eigenstate $\vert E_{n}\rangle$ in Figure~\ref{Fig2}e. When $k_{B}T_{a}/\hbar\omega_{0}=k_{B}T_{\sigma}/\hbar\omega_{0}=2$, namely, the temperatures of the two IHBs are the same, we find that the steady-state populations $p_{n}$ are the same as those of the thermal state (the first group bars numbered from the left) with the same temperature as the two IHBs. This is a signature of thermalization. However, when the temperatures of the two IHBs are different from each other, for example $k_{B}T_{a}/\hbar\omega_{0}=1.5$ and $k_{B}T_{\sigma}/\hbar\omega_{0}=2.5$, the steady-state populations $p_{n}$ always differ from those for the thermal state.

We emphasize that the configuration of the system-bath couplings is very important for the thermalization. We have shown that, when the temperatures of the two baths are different, the JC system   cannot be thermalized. The results will be changed when one of the system-bath couplings is turned off.
To address this point, in Figure~\ref{Fig3}, we show the scaled effective temperatures $k_{B}T_\text{eff}(\omega_{m,n})/\hbar\omega_{0}$ as functions of the energy level indexes $m$ and $n$. Specifically, Figures~\ref{Fig3}a and~\ref{Fig3}b show the results when the bosonic-field bath is decoupled [Figure~\ref{Fig3}a] and the bath of the bosonic field is at zero temperature [Figure~\ref{Fig3}b], respectively. We find that when only one IHB is coupled to the system, the system can be thermalized. However, when the two IHBs are both coupled to the system at different temperatures, the system cannot be thermalized though one of the two IHBs is at zero temperature. This observation also works in the case where the bath of the TLS is decoupled from the system or at zero temperature (see Figures~\ref{Fig3}c and~\ref{Fig3}d). The JC system can be thermalized in the coupling of one IHB case and cannot be thermalized in the nonequilibrium (two different temperatures) two-IHB case. We note that quantum thermalization of the JC model in the single-cavity-bath case has been studied in Ref.~\cite{bib8}. In Figures~\ref{Fig3}e, f, we plot the steady-state populations $p_{n}$ of the eigenstates $\vert E_{n}\rangle$ corresponding to the cases in Figures~\ref{Fig3}a-d. We can see that when the bath of the TLS (bosonic mode) is decoupled, the steady-state populations of the system are the same as those of the thermal state at the same temperature as the coupled bath. While the steady-state populations differ from the thermal-state populations when the temperatures of the two baths are different. It indicates that the JC system can be thermalized when only one bath is coupled while it cannot be thermalized when the coupled two IHBs are at different temperatures, even when one of the two baths is at zero temperature.

The thermalization result can also be evaluated by calculating the trace distance between the steady-state density matrix $\rho_{\text{ss}}$ and the thermal state density matrix $\rho_{\text{th}}$ corresponding to the JC model. The trace distance\textsuperscript{\cite{bib55p5}} between these two density matrices is defined by
$D\left(\rho_{\text{ss}},\rho_{\text{th}}\right) \equiv \frac{1}{2}\Vert\rho_{\text{ss}}-\rho _{\text{th}}\Vert_{1}$, where we introduce the trace norm for an operator $A$ as $\Vert A\Vert_{1} \equiv \text{Tr}[\sqrt{A^{\dagger}A}]$.

In the steady state of the system, the non-diagonal matrix elements of the density matrix $\rho_{\text{ss}}$ expressed in the eigenstate representation are $0$, that is, $\langle E_{n}|\rho_\text{ss}|E_{m}\rangle=0$ for $m\neq n$, then the steady-state density matrix $\rho _{\text{ss}}$ of the JC system can be expressed as
$\rho_{\text{ss}}=\sum_{n=1}^{\infty}p_{n}^{\text{ss}}\left\vert E_{n}\right\rangle\left\langle E_{n}\right\vert$,
where $p_{n}^{\text{ss}}$ denotes the probability corresponding to the eigenstate $\vert E _{n}\rangle$.

When the JC system is in a thermal equilibrium at the inverse temperature $\beta$, the density matrix $\rho_{\text{th}}$ can be written in the eigenstate representation as
$\rho _{\text{th}}=\sum_{n=1}^{\infty}p_{n}^{\text{th}}\left\vert E_{n}\right\rangle\left\langle E _{n}\right\vert$,
where $p_{n}^{\text{th}}=Z_{\text{JC}}^{-1}e^{-\beta E_{n}}$ denotes the thermal-state probability corresponding to the eigenstate $\vert E _{n}\rangle$. The trace distance between the steady-state and thermal-state density matrices can be obtained as
\begin{equation}
D\left(\rho_{\text{ss}},\rho_{\text{th}}\right)=\frac{1}{2}\left\Vert\sum_{n=1}^{\infty}\left(
p_{n}^{\text{ss}}-p_{n}^{\text{th}}\right) \left\vert E_{n}\right\rangle\left\langle E _{n}\right\vert \right\Vert_{1}.
\end{equation}
%%%%%%%%%%%%%%%%%%%%%
\begin{figure}
\center
\includegraphics[width=0.47 \textwidth]{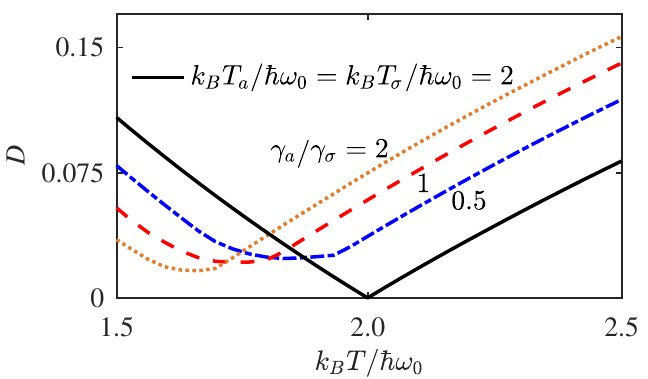}
\caption{\label{Fig4}The trace distance $D$ between the steady-state density matrix of the JC system and the referenced thermal state at the temperature $T$ as a function of the referenced temperature $k_{B}T/\hbar\omega_{0}$ in various cases: $k_{B}T_{a}/\hbar\omega_{0}=k_{B}T_{\sigma}/\hbar\omega_{0}=2$ and $\gamma_{\sigma}/\omega_{0}=\gamma_{a}/\omega_{0}=0.0001$ (black solid line); $k_{B}T_{a}/\hbar\omega_{0}=1.5$, $k_{B}T_{\sigma}/\hbar\omega_{0}=2.5$, $\gamma_{\sigma}/\omega_{0}=0.0001$, and $\gamma_{a}/\gamma_{\sigma}=0.5$ (blue dash-dotted line), $1$ (red dashed line), and $2$ (orange dotted line).}
\end{figure}
%%%%%%%%%%%%%%%%%%%%%

Below we check the trace distance $D$ between the steady state of the open JC system and a referenced thermal state. In the IHB case, when the temperatures of the two baths are different, the temperature of the referenced thermal state is taken as a value between the two bath temperatures. If the trace distance $D$ is zero for one of the value of the referenced temperature, it means that the open JC system is thermalized into a thermal equilibrium state at the referenced temperature. In Figure~\ref{Fig4}, we plot the trace distance $D$ between the steady-state density matrix and a referenced thermal state as a function of the referenced temperature $k_{B}T/\hbar\omega_{0}$ in various cases. We find $D>0$ when the two independent heat baths have different temperatures (the colored lines), which means that the steady-state density matrix $\rho_{\text{ss}}$ of the JC system is not the density matrix $\rho_{\text{th}}$ of a thermal state, that is, the open JC model cannot be thermalized when the two independent heat baths have different temperatures. For comparison, we also plot the trace distance $D$ (the black solid line) when the two independent heat baths have the same temperature $k_{B}T_{a}/\hbar\omega_{0}=k_{B}T_{\sigma}/\hbar\omega_{0}=2$. Here, we see $D=0$ at temperature $k_{B}T/\hbar\omega_{0}=2$ and $D>0$ for $k_{B}T/\hbar\omega_{0}\neq2$. The relation $D=0$ means that the steady-state density matrix $\rho_{\text{ss}}$ is a thermal state density matrix at the same temperature as the bath temperature, which implies that the JC model can be thermalized when the two independent heat baths have the same temperatures. Based on the above analyses, we can summarize that the JC system can be thermalized in three special cases: The two baths have the same temperatures or only one of the two subsystems is coupled to a heat bath.

To clarify the physical mechanism of the thermalization in this open JC system, we need to analyze the steady state of the system, which is determined by the stationary solution of quantum master equation~(\ref{master-equt}). From the master equation, we can analyze the jump operators and the physical processes induced by the system-bath interaction. In the system-bath interactions, we expand the system operators $\sigma_{x}$ and $(a+a^{\dagger})$ with the eigenstates of the JC Hamiltonian. As a result, we should analyze the transitions induced by the system-bath interactions in the eigenstate representation of the JC Hamiltonian.
In equation~(\ref{master-equt}), we see that there are many transition channels between two eigenstates in neighboring subspaces with $n$- and $(n+1)$-excitations. In particular, these dissipators corresponding to all these transition channels take the form as the Lindblad form. The Lindblad dissipators have the structure of quantum dynamical semigroup and the system has the property of being ergodic. For a pair of transition processes [for example $\mathcal{D}[\vert\varepsilon_{n,\beta_{n}}\rangle\langle\varepsilon_{n+1,\alpha_{n+1}}\vert]\tilde{\rho}(t)$ and $\mathcal{D}[\vert\varepsilon_{n+1,\alpha_{n+1}}\rangle\langle\varepsilon_{n,\beta_{n}}\vert]\tilde{\rho}(t)$ for the downward and upward transitions], the effective transition rates are given by
$\sum_{l=\sigma,a}\frac{1}{2}\gamma_{l}(\omega_{n,\alpha_{n+1},\beta_{n}})\vert\chi_{l,\alpha_{n+1},\beta_{n}}\vert^{2}[\bar{n}_{l}(\omega_{n,\alpha_{n+1},\beta_{n}})+1]$ and $\sum_{l=\sigma,a}\frac{1}{2}\gamma_{l}(\omega_{n,\alpha_{n+1},\beta_{n}})\vert\chi_{l,\alpha_{n+1},\beta_{n}}\vert^{2}\bar{n}_{l}(\omega_{n,\alpha_{n+1},\beta_{n}})$.
In principle, we can always define an effective temperature corresponding to the steady state populations of the two eigenstates $\vert\varepsilon_{n,\beta_{n}}\rangle$ and $\vert\varepsilon_{n+1,\alpha_{n+1}}\rangle$ based on the downward and upward transitions rates (using the detailed balance condition). However, when the two baths have different temperatures, the temperatures of each pairs of two states are different and hence the state of the JC system cannot be expressed as a thermal state density matrix. Only in the above mentioned three special cases, the temperatures corresponding to all these pairs of two states are the same and then the JC system can be thermalized. Physically, the thermalization is a consequence of the Kubo-Martin-Schwinger (KMS) condition, which is satisfied in these three special cases.\textsuperscript{\cite{bib17}} Note that the thermal equilibrium in the JC model have been analyzed in the case of single heat bath for the bosonic mode using the Heisenberg-picture operator method.\textsuperscript{\cite{bib56}}

\section{Quantum thermalization in the CHB case  \label{Sec:CHB}}

In this section, we study quantum thermalization of the JC model in the CHB case, in which the TLS and the bosonic mode are immersed in a common heat bath. In the CHB case, the evolution of the JC system is governed by the following quantum master equation
\begin{equation}
\frac{d}{dt}\rho_{S}(t)=-\frac{i}{\hbar}[H_{S},\rho_{S}(t)]+L_{\text{CHB}}[\rho_{S}(t)], \label{MasteqCHB}
\end{equation}
where the dissipator $L_{\text{CHB}}[\rho_{S}(t)]$ is given by
\begin{eqnarray}
&&L_{\text{CHB}}[\rho_{S}(t)]\nonumber\\
&=&\sum_{n=0}^{\infty}\sum_{\alpha_{n},\beta_{n}}\sum_{l=\sigma,a,X}\frac{1}{2}\gamma_{l}(\omega_{n,\alpha_{n+1},\beta_{n}})\vert\chi_{l,\alpha_{n+1},\beta_{n}}\vert^{2}\nonumber\\
&&\times[\bar{n}_{l}(\omega_{n,\alpha_{n+1},\beta_{n}})+1]\mathcal{D}[\vert\varepsilon_{n,\beta_{n}}\rangle\langle\varepsilon_{n+1,\alpha_{n+1}}\vert]\rho_{S}(t) \nonumber\\
&&+\sum_{n=0}^{\infty}\sum_{\alpha_{n},\beta_{n}}\sum_{l=\sigma,a,X}\frac{1}{2}\gamma_{l}(\omega_{n,\alpha_{n+1},\beta_{n}})\vert\chi_{l,\alpha_{n+1},\beta_{n}}\vert^{2}\nonumber\\
&&\times\bar{n}_{l}(\omega_{n,\alpha_{n+1},\beta_{n}})\mathcal{D}[\vert\varepsilon_{n+1,\alpha_{n+1}}\rangle\langle\varepsilon_{n,\beta_{n}}\vert]\rho_{S}(t)\label{master-cavity2},
\end{eqnarray}
with the Lindblad superoperator $\mathcal{D}[O]$ defined in Sec.~\ref{Sec:IHB}.
The decay rates in equation~(\ref{master-cavity2}) are defined by
\begin{subequations}
\begin{align}
\gamma_{\sigma}(\omega_{n,\alpha_{n+1},\beta_{n}})=&2\pi\varrho_{c}(\omega_{n,\alpha_{n+1},\beta_{n}})\lambda^{2}(\omega_{n,\alpha_{n+1},\beta_{n}}),\\
\gamma_{a}(\omega_{n,\alpha_{n+1},\beta_{n}})=&2\pi\varrho_{c}(\omega_{n,\alpha_{n+1},\beta_{n}})\eta^{2}(\omega_{n,\alpha_{n+1},\beta_{n}}),\\
\gamma_{X}(\omega_{n,\alpha_{n+1},\beta_{n}})=&\sqrt{\gamma_{\sigma}(\omega_{n,\alpha_{n+1},\beta_{n}})\gamma_{a}(\omega_{n,\alpha_{n+1},\beta_{n}})},
\end{align}
\end{subequations}
which correspond to the dissipations through the TLS, the bosonic mode, and the cross effect between the two subsystems, respectively. Note that the variable $\varrho_{c}$ is the density of state of the common heat bath.
The parameters $\bar{n}_{\sigma}(\omega_{n,\alpha_{n+1},\beta_{n}})=\bar{n}_{a}(\omega_{n,\alpha_{n+1},\beta_{n}})=\bar{n}_{X}(\omega_{n,\alpha_{n+1},\beta_{n}})=[\exp(\hbar\omega_{n,\alpha_{n+1},\beta_{n}}/k_{B}T)-1]^{-1}$ are the average thermal excitation numbers at the bath temperature $T$. The transition coefficients $\chi_{\sigma,\alpha_{n+1},\beta_{n}}$ and $\chi_{a,\alpha_{n+1},\beta_{n}}$ have been defined in Sec.~\ref{Sec:IHB}, and the cross transition coefficient is defined by $\chi_{X,\alpha_{n+1},\beta_{n}}=\sqrt{2\chi_{\sigma,n,\alpha_{n+1}}\chi_{a,n,\alpha_{n+1}}}$.
Based on quantum master equation~(\ref{MasteqCHB}), we analyze the temperatures associated with these eigenstates, and find that these temperatures are the same as that of the CHB, which means that the JC model can approach a thermal equilibrium with the CHB, namely the JC model can be thermalized in the CHB case. Note that this is a consequence of detailed balance condition under the relation $\bar{n}_{\sigma}=\bar{n}_{a}=\bar{n}_{X}$.

\section{Vanishing thermal entanglement in the thermalized JC model \label{thementanglement}}

In the above section, we have shown that the JC system can be thermalized in three special IHB cases. When the JC system is thermalized at the temperature $T$, its density matrix can be written as $\rho_{\text{th}}(T)=Z_{\text{JC}}^{-1}\exp(-\beta H_\text{JC})$ with $\beta=1/(k_{B}T)$. Intuitively, the thermal entanglement should exist in the thermalized JC system. The excited eigenstates $\vert\varepsilon_{n\pm}\rangle$ $(n\geq 1)$ of the JC model are entangled states, due to the interaction between the TLS and the bosonic mode. In the finite-temperature cases, these eigenstates $\vert\varepsilon_{n\pm}\rangle$ $(n\geq 1)$ will be occupied. Then the TLS and the bosonic mode will be entangled with each other in the thermalized states. Below we study the steady-state quantum entanglement between the TLS and the bosonic field by calculating the logarithmic negativity under various system parameters. We find and show that the thermal entanglement is very small in the JC parameter regime.

For the JC system in the density matrix $\rho$, the logarithmic negativity\textsuperscript{\cite{bib57,bib58}} can be calculated by
\begin{equation}
N=\log_{2}{\Vert\rho^{T_{\sigma}}\Vert_{1}},\label{NEG}
\end{equation}
where``$T_{\sigma}$'' denotes the partial transpose with respect to the TLS. Below we calculate the quantum entanglement of the thermalized JC system, which is in the thermal state $\rho_{\text{th}}(T)=Z_{\text{JC}}^{-1}\exp(-\beta H_{\text{JC}})$, where the partition function can be calculated as
$Z_{\text{JC}}=2\sum_{n=1}^{\infty}\exp[-\hbar\beta\omega_{c}(n-1/2)]\cosh(\hbar\beta\Omega_{n}/2)+e^{\beta\hbar\omega_{0}/2}$.
In this case, the quantum entanglement between the TLS and the bosonic field can be analytically calculated based on Eq.~(\ref{NEG}).

When the JC system is in the thermal state $\rho_{\text{th}}(T)$, the corresponding probabilities of these eigenstates $|E_{1}\rangle=|\varepsilon_{0,0}\rangle$, $|E_{2n}\rangle=|\varepsilon_{n-}\rangle$, and $|E_{2n+1}\rangle=|\varepsilon_{n+}\rangle$ are given by $p_{1}=Z_{\text{JC}}^{-1}e^{-\beta E_{1}}$, $p_{2n}=Z_{\text{JC}}^{-1}e^{-\beta E_{2n}}$, $p_{2n+1}=Z_{\text{JC}}^{-1}e^{-\beta E_{2n+1}},\hspace{0.5cm} n=1,2,3,\cdots$, where $E_{1}=-\hbar\omega_{0}/2$, $E_{2n}=\hbar\omega_{c}(n-1/2)-\hbar \Omega _{n}/2$, and $E_{2n+1}=\hbar \omega_{c}(n-1/2)+\hbar \Omega _{n}/2$ for $n\geq 1$.
To calculate the logarithmic negativity of the thermal state, we express the density matrix of the thermal state using the bare states as
\begin{eqnarray}
\rho _{\text{th}}(T)&=&A_{0}|g\rangle_{q}|0\rangle_{c}\langle g|_{q}\langle 0|_{c}
+\sum_{n=1}^{\infty}\left[A_{n}|g\rangle_{q}|n\rangle _{c}\langle g|_{q}\langle n|_{c} \right.\nonumber\\
&&\left.+B_{n}(|g\rangle_{q}|n\rangle_{c}\langle e|_{q}\langle n-1|_{c}+\text{H.c.}) \right.\nonumber\\
&&\left.+C_{n}|e\rangle _{q}|n-1\rangle _{c}\langle e|_{q}\langle n-1|_{c}\right],\label{loth}
\end{eqnarray}
where we introduce the variables $A_{0}=p_{1}$ and
\begin{subequations}
\begin{align}
A_{n}=&\sin ^{2}(\theta _{n}/2)p_{2n+1}+\cos ^{2}(\theta_{n}/2)p_{2n},\\
B_{n}=&\sin(\theta_{n}/2)\cos(\theta_{n}/2)(p_{2n+1}-p_{2n}),\\
C_{n}=&\cos^{2}(\theta_{n}/2)p_{2n+1}+\sin ^{2}(\theta_{n}/2)p_{2n},
\end{align}
\end{subequations}
for $n=1,2,3,\cdots$. Based on Eq.~(\ref{loth}), we can obtain the following relation
\begin{eqnarray}
M&=&(\rho_{\text{th}}^{T\sigma})^{\dagger }\rho_{\text{th}}^{T\sigma}=C_{1}^{2}|e\rangle_{q}|0\rangle_{c}\langle e|_{q}\langle 0|_{c} \nonumber\\
&&+\sum_{n=0}^{\infty}\left[(A_{n}^{2}+B_{n+1}^{2})|g\rangle_{q}|n\rangle_{c}\langle g|_{q}\langle n|_{c}\right.\nonumber\\
&&\left.+(B_{n+1}^{2}+C_{n+2}^{2})|e\rangle _{q}|n+1\rangle _{c}\langle e|_{q}\langle n+1|_{c}\right.\nonumber \\
&&\left.+B_{n+1}(A_{n}+C_{n+2})(|g\rangle_{q}|n\rangle _{c}\langle e|_{q}\langle n+1|_{c}+\text{H.c.})\right].
\end{eqnarray}
If we express the operator $M$ with these basis states ordered by $\{|e,0\rangle, |g,0\rangle, |e,1\rangle, |g,1\rangle, |e,2\rangle, \cdots,|g,n\rangle, |e,n+1\rangle, \cdots\}$, then the operator $M$ can be decomposed into a direct sum of a one-dimensional matrix $C_{1}^{2} $ and an infinite number of $2\times 2$ matrices, i.e., $M=C_{1}^{2}\oplus M^{[1]}\oplus M^{[2]}\oplus\cdots\oplus M^{[n+1]}\oplus\cdots$. Here, the $(n+1)$th ($n\geq0$) $2\times 2$ matrix is associated with the subspace defined with the basis states $\{|g,n\rangle,|e,n+1\rangle\}$ and it can be written as
\begin{equation}
M^{[n+1]}=\left(\begin{array}{cc}
A_{n}^{2}+B_{n+1}^{2} & B_{n+1}(A_{n}+C_{n+2}) \\
B_{n+1}(A_{n}+C_{n+2}) & B_{n+1}^{2}+C_{n+2}^{2}\label{matrix}
\end{array}
\right).
\end{equation}
The trace norm $\Vert\rho_{\text{th}}^{T\sigma}\Vert_{1}$ can then be obtained by calculating the eigenvalues of the matrix  $M^{[n+1]}$ in Eq.~(\ref{matrix}). For below convenience, we denote $\lambda _{n+1,1}$ and $\lambda_{n+1,2}$ as the  eigenvalues of the matrix $M^{[n+1]}$. Then the trace norm can be obtained as
\begin{equation}
\Vert\rho_{\text{th}}^{T\sigma}\Vert_{1}=C_{1}+\sum_{n=0}^{\infty }\left(\sqrt{\lambda_{n+1,1}}+\sqrt{\lambda_{n+1,2}}\right),    \label{rhotsigma}
\end{equation}
which indicates that the expression of $\Vert\rho_{\text{th}}^{T\sigma}\Vert_{1}$ can be obtained analytically by calculating the eigenvalues of the matrices $M^{[n+1]}$. Based on Eqs.~(\ref{NEG}) and~(\ref{rhotsigma}), the logarithmic negativity of the thermal state can be calculated.

In this work, we consider the case where the JC Hamiltonian is obtained by making the rotating-wave approximation (RWA) in the quantum Rabi model describing the dipole interaction between a two-level atom and a single-mode cavity field. According to the parameter space of cavity-QED systems, the RWA is valid in both the weak-coupling and strong-coupling regimes, in which the atom-field coupling strength is smaller and larger than the decay rates, respectively. However, when the atom-field coupling strength reaches an appreciable fraction of the resonance frequencies of the subsystems (atom and field), the system enters the ultrastrong coupling regime, in which the RWA is not valid. Based on the recent studies in the field of ultrastrong coupling, people usually takes this bound value as $g/\omega_{0}=1/s=0.1$.\textsuperscript{\cite{bib60,bib61}} When $g/\omega_{0}>0.1$ ($s<10$), the parameter regime is referred as the ultrastrong-coupling regime,\textsuperscript{\cite{bib60,bib61}} in which the RWA\textsuperscript{\cite{bib1}} is no longer justified.
%%%%%%%%%%%%%%%%%%%%%
\begin{figure}
\center
\includegraphics[width=0.47 \textwidth]{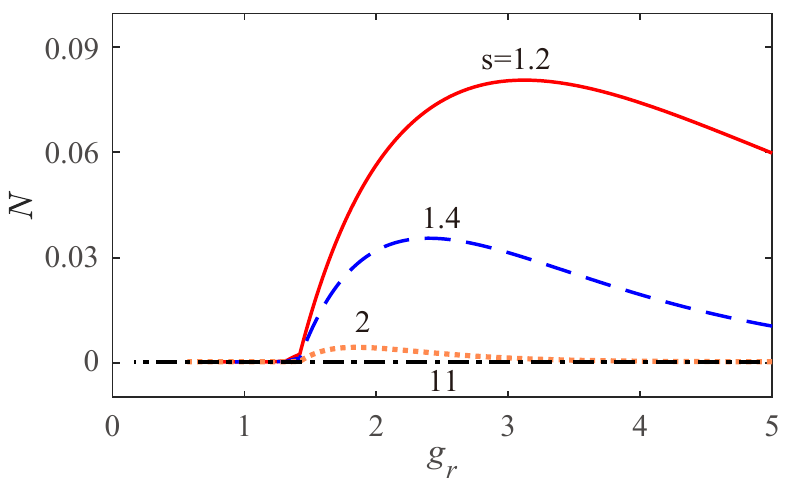}
\caption{\label{Fig5} The logarithmic negativity $N$ of the JC system in the thermal state versus $g_{r}=\hbar g\beta$ when $s=\omega_{0}/g$ takes various values: $s=1.2$, $1.4$, $2$, and $11$. Other parameter used is $\delta=0$.}
\end{figure}
%%%%%%%%%%%%%%%%%%%%%%%

In Figure~\ref{Fig5}, we plot the logarithmic negativity describing the thermal entanglement of the JC model as a function of $g_{r}=\hbar g\beta$ in the resonant case $\delta=0$ when $s=\omega_{0}/g$ takes various values: $s=1.2$, $1.4$, $2$, and $11$. Here we should point out that for the JC model, a valid parameter condition should be $s>10$ such that the RWA can be justified. It can be seen from Figure~\ref{Fig5} that for a given $s$, there is an optimal value range of $g_{r}$ corresponding to a large thermal entanglement. On one hand, the JC system will be in its ground state in the zero-temperature case and hence these is no quantum entanglement in this case. On the other hand, in the high-temperature limit, the thermal noise will destroy quantum effect and then the quantum entanglement will disappear in this case. For a given value $g_{r}$, we find that the logarithmic negativity is larger for a smaller value of $s$. This is because quantum entanglement will increase with the coupling between the TLS and the bosonic mode. When the coupled system enters the JC parameter regime ($s>10$), the entanglement disappears gradually, which means the vanishing thermal entanglement in the JC model.

To show the phenomenon of vanishing thermal entanglement in the JC system, in the following we present an analytical verification of this result. Mathematically, we find that, when $A_{n}C_{n+2}-B_{n+1}^{2}\geq 0$, we have the relation
\begin{equation}
\sqrt{\lambda _{n+1,1}}+\sqrt{\lambda _{n+1,2}}=A_{n}+C_{n+2}.\label{gret0cond}
\end{equation}
Moreover, when $A_{n}C_{n+2}-B_{n+1}^{2}<0$, we have
\begin{equation}
\sqrt{\lambda_{n+1,1}}+\sqrt{\lambda_{n+1,2}}=\sqrt{(A_{n}-C_{n+2})^{2}+4B_{n+1}^{2}}.\label{smal0cond}
\end{equation}
In principle, we can obtain the logarithmic negativity based on Eqs.~(\ref{NEG}) and~(\ref{rhotsigma}). Below, we consider the resonant JC Hamiltonian case for simplicity.

\begin{figure}
\center
\includegraphics[width=0.47 \textwidth]{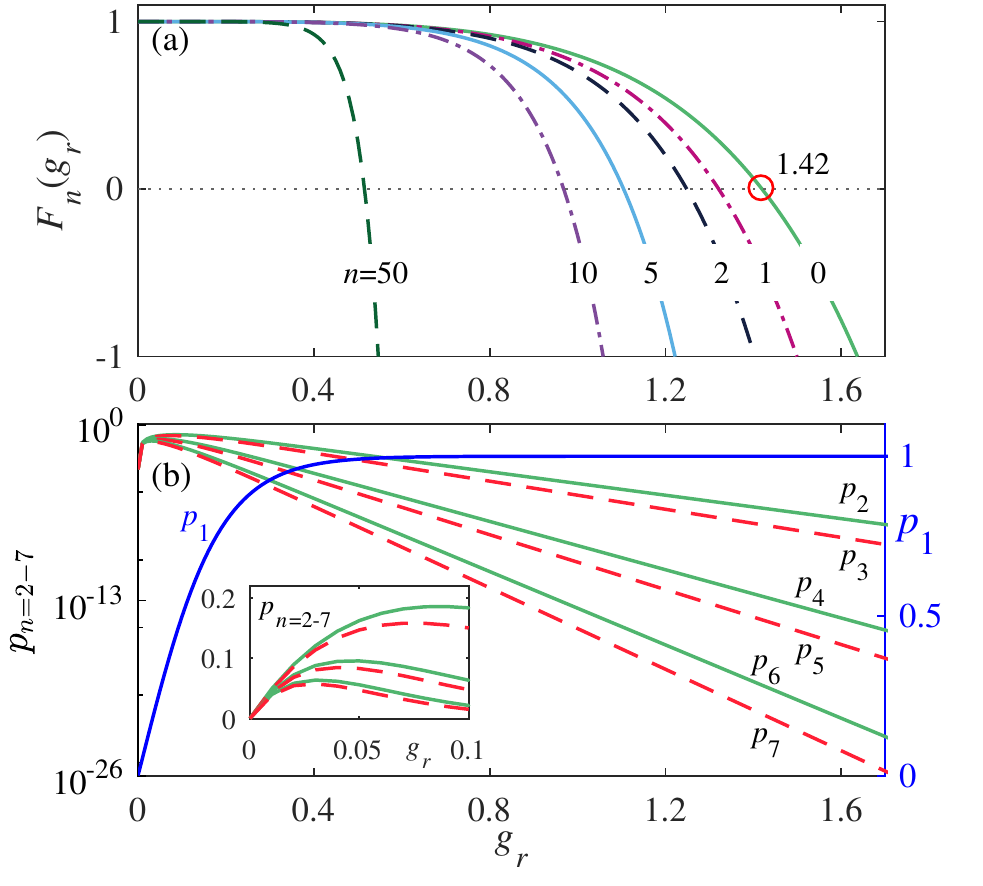}
\caption{ \label{Fig6}a) $F_{n}(g_{r})$ as a function of $g_{r}$ at various $n$. b) The probabilities $p_{n}$ of the eigenstate $|E_{n}\rangle$ as functions of $g_{r}$ when the JC system is in the thermal state. Other parameter used is $s=11$.}
\end{figure}

In the resonance case $\delta=0$, we introduce the following function to characterize the sign of the conditional variable  $F_{n}(g_{r})\equiv A_{n}C_{n+2}-B_{n+1}^{2}$,
\begin{eqnarray}
F_{n}=\cosh(\sqrt{n}g_{r})\cosh(\sqrt{n+2}g_{r})-\sinh^{2}(\sqrt{n+1}g_{r}), \label{condition}
\end{eqnarray}
where we introduce the relative coupling strength $g_{r}=\hbar g\beta=\hbar g/k_{B}T$. Obviously, $F_{n}(g_{r})$ is a function of $n$ and $g_{r}$. Below, we discuss the behavior of the function $F_{n}(g_{r})$ when $n$ and $g_{r}$ take different values. In Figure~\ref{Fig6}a, we plot $F_{n}(g_{r})$ as a function of $g_{r}$ when $n$ takes various values. It can be seen that the curves move left with the increase of $n$ and $F_{n}(g_{r})$ decreases monotonically as $g_{r}$ increases, and that different values of $n$ correspond to different values of $g_{r}$ satisfying $F_{n}(g_{r})=0$. For a given $n$, we obtain a critical value of $g^{n}_{rc}$, which is located at the cross point between the curve $F_{n}(g_{r})$ and the curve $F_{n}(g^{n}_{rc})=0$. In particular, for $n=0$ we have the critical value $g^{0}_{rc}\approx1.42$. For a lager $n$, a smaller value of $g^{n}_{rc}$ can be obtained because the curves move left with the increase of $n$. The values of these critical points are very important because these values determine that which of the relations given in Eqs.~(\ref{gret0cond}) and~(\ref{smal0cond}) should be used to calculate the value of $\sqrt{\lambda _{n+1,1}}+\sqrt{\lambda _{n+1,2}}$.
%%%%%%%%%%%%%%%%%%%%%
%%%%%%%%%%%%%%%%%%%%%%%

Below, we first analyze two special cases of the coupling strength $g_{r}$.

(i) The decoupling case: $g=0$. In this case,
the inequality $A_{n}C_{n+2}-B_{n+1}^{2}\geq 0$ holds for all $n$ ($n=0$, $1$, $2$, $\cdots $), then we can calculate the trace norm with Eq.~(\ref{gret0cond}) as
\begin{eqnarray}
\Vert\rho_{\text{th}}^{T\sigma}\Vert_{1}&=&C_{1}+\sum_{n=0}^{\infty}(A_{n}+C_{n+2})=\sum_{n=1}^{\infty} p_{n}=1,
\end{eqnarray}
and then the logarithmic negativity $N(\rho_{\text{th}})=\log_{2}\Vert\rho_{\text{th}}^{T\sigma}\Vert_{1}=0$,
which is a straightforward result: there is no thermal entanglement in the decoupling case.

(ii) The case of $g_{r}\geq1.42$. In this case, $F_{n}(g_{r})<0$ for all $n$. Then we can calculate the trace norm with Eq.~(\ref{smal0cond}) as
\begin{eqnarray}
\Vert\rho_{\text{th}}^{T\sigma}\Vert_{1}&=&\frac{(p_{3}+p_{2})}{2}+\frac{1}{2}\{[2p_{1}-(p_{5}+p_{4})]^{2}+4(p_{3}-p_{2})^{2}\}^{\frac{1}{2}}\nonumber\\
&&+\sum_{n=1}^{\infty }\frac{1}{2}\{[(p_{2n+1}+p_{2n})-(p_{2n+5}+p_{2n+4})]^{2}\nonumber\\
&&+4(p_{2n+3}-p_{2n+2})^{2}\}^{\frac{1}{2}}.
\label{eqbthtranormgrt1p42}
\end{eqnarray}
The above expression can be simplified by analyzing the probability distributions $p_{n}$ of these eigenstates of the JC model. By denoting $s=\omega_{0}/g$, then the probability of the ground state can be written as $p_{1}=[1+\sum_{n=1}^{\infty }e^{-g_{r}sn}(e^{g_{r}\sqrt{n}}+e^{-g_{r}\sqrt{n}})]^{-1}$, which satisfies the relation $[1+1/(e^{g_{r}(s-1)}-1)+1/(e^{g_{r}s}-1)]^{-1}<p_{1}<[1+2/(e^{g_{r}s}-1)]^{-1}$. In the JC regime, we take $s>10$ in the following estimations. For example, when we take $s=11$, we find that $p_{1}\approx1$ and $p_{n>1}\approx0$ for $g_{r}>1.42$ (Figure~\ref{Fig6}b), then we have $\Vert\rho_{\text{th}}^{T\sigma}\Vert_{1}\approx1$ according to Eq.~(\ref{eqbthtranormgrt1p42}). This result is understandable by analyzing the relation between the thermal excitation energy the transition energy between the ground state and the first excited eigenstate. When $s>10$ and $g_{r}>1.42$, we have $\hbar \omega _{0}>14.2k_{B}T$. Then $E_{2}-E_{1}=$ $\hbar \omega _{0}-\hbar g=g_{r}(s-1)k_{B}T>12.78k_{B}T$, which means that the transition probability excited by the thermal noise is negligible. The JC system will stay in its ground state $\vert g,0\rangle$ and hence there is no thermal entanglement.
%%%%%%%%%%%%%%%%%%%%%%%%%%%
\begin{center}
\begin{table*}
\label{table2}
\caption{This table shows the values of the involved excitation number $n_{0}+2$ and the truncation dimension parameter $n_{\text{max}}$ when the relative coupling strength $g_{r}$ takes various values. For a given value of $g_{r}$, the $n_{0}$ is determined by the relations $F_{n_{0}}(g_{r})\geq 0$ and $F_{n_{0}+1}(g_{r})< 0$, and the truncation dimension parameter $n_{\text{max}}$ is determined by the probability threshold  $p_{n}< 10^{-20}$. The parameter $s=11$ is chosen such that the RWA is justified in the JC parameter regime.}
%\begin{ruledtabular}
\centering
\footnotesize
\setlength{\tabcolsep}{4mm}{
\begin{tabular}{cccccccccc}
\hline
\hline
  $g_{r}$ &$0.1$ &$0.2$ &$0.3$&$0.4$ &$0.6$ &$0.8$ &$1.0$ &$1.2$&$1.4$\\
  \midrule
  $n_{0}+2$&$8731$&$1467 $&$489 $&$216 $&$63 $&$24 $&$10 $&$4$&$2 $\\
  $n_{\text{max}}$&$40$ &$21$ &$12$ &$10$ &$6 $&$5$ & $4$&$3$ &$2$\\
\hline
\hline
\end{tabular}}
%\end{ruledtabular}
\end{table*}
\end{center}
%%%%%%%%%%%%%%%%%%%%%%

We now consider a general case, namely the coupling strength $0<g_{r}<1.42$. In this case, there always exists an $n_{0}$ such that $F_{n_{0}}(g_{r})\geq 0$ and $F_{n_{0}+1}(g_{r})< 0$. Note that the contributions corresponding to the terms $n>n_{0}$ and $n\leq n_{0}$ should be calculated with the formula~(\ref{gret0cond}) and~(\ref{smal0cond}), respectively. Then according to Eqs.~(\ref{rhotsigma}), (\ref{gret0cond}), and~(\ref{smal0cond}), we have
\begin{eqnarray}
\Vert\rho_{\text{th}}^{T\sigma }\Vert_{1}&=&\sum_{n=1}^{2n_{0}+1}p_{n}+\frac{1}{2}(p_{2n_{0}+5}+p_{2n_{0}+4}+p_{2n_{0}+3}+p_{2n_{0}+2}) \nonumber\\
&&+\sum_{n=n_{0}+1}^{\infty }\frac{1}{2}\left\{[(p_{2n+1}+p_{2n})-(p_{2n+5}+p_{2n+4})]^{2}\right. \nonumber\\
&&\left.+4(p_{2n+3}-p_{2n+2})^{2}\right\}^{\frac{1}{2}}.
\end{eqnarray}
In a realistic calculation, we need to truncate the Hilbert space of the JC system up to an excitation number, denoting as $n_{\text{max}}$. The value of $n_{\text{max}}$ is determined by setting a threshold value of the population $p_{n_{\text{max}}}$ for the truncation. In our simulation, we choose $p_{n_{\text{max}}}=10^{-20}$. This means that we treat $p_{n>n_{\text{max}}}\approx0$ in the calculation of the logarithmic negativity. In addition, for a given $g_{r}$, the value of $n_{0}$ can be determined according to the relations $F_{n_{0}}(g_{r})\geq 0$ and $F_{n_{0}+1}(g_{r})< 0$. Here, the value of $n_{0}$ determines the number of these $2\times2$ matrices which should be taken into account in $M$. For the given $n_{0}$, the associated $M^{[n_{0}+1]}$ matrix involves the basis states $|g,n_{0}\rangle$ and $|e,n_{0}+1\rangle$. Here $|e,n_{0}+1\rangle$ is a basis state in the subspace with the excitation number $n_{0}+2$. For the JC model, we can determine the values of the two parameters $n_{\text{max}}$ and $n_{0}$, as shown in Table $1$. Here, we can see that $n_{0}+2\geq n_{\text{max}}$, which is also satisfied $s>11$. Therefore, all the conditions $F_{n\leq (n_{\text{max}}-2)}(g_{r})\geq 0$ are satisfied, and the value of the trace norm can be calculated within the truncated subspace as
\begin{eqnarray}
\left\Vert\rho_{\text{th}}^{T\sigma}\right\Vert_{1} &=&\frac{1}{2}\left\{\sum_{n=1}^{2n_{\text{max}}+1}2p_{n}-(p_{2n_{\text{max}}-1}+p_{2n_{\text{max}}-2})\right.\nonumber\\
&&+\left.\left[(p_{2n_{\text{max}}-1}+p_{2n_{\text{max}}-2})^{2}\right.\right.\nonumber\\
&&\left.\left.+(p_{2n_{\text{max}}+1}-p_{2n_{\text{max}}})^{2}\right]^{1/2}\right\}.
\end{eqnarray}
As the probabilities around the truncation dimension parameter $n_{\text{max}}$ is negligible, we can obtain approximately the trace norm as $\Vert\rho_{\text{th}}^{T\sigma }\Vert_{1}\simeq\sum_{n=1}^{2n_{\text{max}}+1}p_{n}= \text{Tr}(\rho_{\text{th}})=1$, and then the logarithmic negativity can be obtained as $N\approx0$.

We note that the vanishing thermal entanglement is a {counterintuitive} phenomenon in this system. In the resonant JC Hamiltonian, the ground state is a separate state, but all other eigenstates are entangled states. In particular, there excited eigenstates take the form of the Bell states (the maximal entangled states in the $2\times2$ Hilbert space) in the corresponding subspaces. Intuitively, if the excited eigenstates have non-negligible populations, considerable thermal entanglement will exist between the TLS and the bosonic mode. In Figure~\ref{Fig6}b, we plot the probabilities of these eigenstates as functions of the parameter $g_{r}$ in the JC regime ($s=11$). Here we can see that these excited eigenstates have considerable populations ($p_{2,3}\approx0.15$, see the inset) in a range of $g_{r}$ (corresponding to a moderate bath temperature). However, the logarithmic negativity corresponding to the thermal state is negligible small ($N<10^{-8}$), as shown in Figure~\ref{Fig5}. It should be mentioned that the correlation between the two subsystems of the JC system in the thermal state has been studied.\textsuperscript{\cite{bib64}} The correlation is also very small in the JC parameter range.

\section{Conclusion   \label{Sec:Conclu}}

In conclusion, we have studied quantum thermalization of the JC system coupled with either two IHBs or a CHB.
We have derived two quantum master equations in the eigenstate representation in these cases. We have analyzed the populations of these dressed states in the steady state and calculated the temperatures associated with these levels. In the IHB case, we have found that, when the two IHBs have the same temperature (different temperatures), the JC system can (cannot) be thermalized. When one of the two IHBs is decoupled from the JC system, then the JC system can be thermalized with the contacted bath. In the CHB case, the system can always be thermalized. In addition, we have studied the thermal entanglement between the TLS and the bosonic field. We have found and proved a phenomenon of vanishing thermal entanglement.

\section*{Acknowledgments}
J.-Q.L. would like to thank Chen Wang and Zhihai Wang for helpful discussions. J.-F.H. is supported in part by the National Natural Science Foundation of China (Grant No. 11505055), Scientific Research Fund of Hunan Provincial Education Department (Grant No. 18A007), and Natural Science Foundation of Hunan Province, China (Grant No. 2020JJ5345).
 J.-Q.L. was supported in part by National Natural Science Foundation of China (Grant Nos. 11822501, 11774087, and 11935006), Natural Science Foundation of Hunan Province, China (Grant No. 2017JJ1021), and Hunan Science and Technology Plan Project (Grant No. 2017XK2018).

\section*{Conflict of interest}
The authors declare no conflicts of interest.

%\bibliographystyle{andp2012}
%\bibliography{RefQT}

\begin{thebibliography}{9}
\bibitem{bib1} \textsc{ E.\,T. Jaynes}, \textsc{F.\,W. Cummings}, Proc. IEEE \textbf{1963}, 51, 89.

\bibitem{bib2} \textsc{P.\,R. Berman}, Cavity Quantum Electrodynamics, Academic Press, Boston \textbf{1994}.

\bibitem{bib3} \textsc{B.\,W. Shore}, \textsc{P.\,L. Knight}, J. Mod. Opt. \textbf{1993}, 40, 1195.

\bibitem{bib4} \textsc{J.\,M. Raimond},  \textsc{M.\,Brune}, \textsc{S.~Haroche}, Rev. Mod. Phys. \textbf{2001}, 73, 565.

\bibitem{bib5} \textsc{S. Haroche}, \textsc{J.-M. Raimond}, Exploring the Quantum: Atoms, Cavities and Photons, Oxford University Press, Oxford \textbf{2006}.

\bibitem{bib6} \textsc{M. Brune}, \textsc{F.\,Schmidt-Kaler}, \textsc{A. Maali}, \textsc{J. Dreyer}, \textsc{E. Hagley}, \textsc{J.\,M. Raimond}, \textsc{S. Haroche},  Phys. Rev. Lett. \textbf{1996}, 76, 1800.

\bibitem{bib7} \textsc{D. Meschede}, \textsc{H. Walther}, \textsc{G. M\"{u}ller},  Phys. Rev. Lett. \textbf{1985}, 54, 551.

\bibitem{bib8} \textsc{Y.\,F. Zhu}, \textsc{D.\,J. Gauthier}, \textsc{S.\,E. Morin}, \textsc{Q.\,L. Wu}, \textsc{H.\,J. Carmichael}, \textsc{T.\,W. Mossberg}, Phys. Rev. Lett. \textbf{1990}, 64, 2499.

\bibitem{bib9} \textsc{R.\,J. Thompson}, \textsc{G. Rempe}, \textsc{H.\,J. Kimble},  Phys. Rev. Lett. \textbf{1992}, 68, 1132.

\bibitem{bib10} \textsc{A. Wallraff}, \textsc{D.\,I. Schuster}, \textsc{A. Blais}, \textsc{L. Frunzio}, \textsc{R.-S. Huang}, \textsc{J. Majer}, \textsc{S. Kumar}, \textsc{S.\,M. Girvin}, \textsc{R.\,J. Schoelkopf},  Nature  \textbf{2004}, 431, 162.

\bibitem{bib11} \textsc{A. Boca}, \textsc{R. Miller}, \textsc{K.\,M. Birnbaum}, \textsc{A.\,D. Boozer}, \textsc{J. McKeever}, \textsc{H.\,J. Kimble},  Phys. Rev. Lett. \textbf{2004}, 93, 233603.

\bibitem{bib12} \textsc{T. Yoshie}, \textsc{A. Scherer}, \textsc{J. Hendrickson}, \textsc{G. Khitrova}, \textsc{H.\,M. Gibbs}, \textsc{G. Rupper}, \textsc{C. Ell}, \textsc{O.\,B. Shchekin}, \textsc{D.\,G. Deppe}, Nature \textbf{2004}, 432, 200.

\bibitem{bib13} \textsc{G. Khitrova}, \textsc{H.\,M. Gibbs}, \textsc{M. Kira}, \textsc{S.\,W. Koch}, \textsc{A. Scherer}, Nat. Phys. \textbf{2006}, 2, 81.

\bibitem{bib14} \textsc{H. Toida}, \textsc{T. Nakajima}, \textsc{S. Komiyama},  Phys. Rev. Lett. \textbf{2013}, 110, 066802.  Phys. Rev. Lett. \textbf{2013}, 111, 249701.


\bibitem{bib15} \textsc{J.\,H. Eberly}, \textsc{N.\,B. Narozhny}, \textsc{J.\,J. Sanchez-Mondragon}, Phys. Rev. Lett. \textbf{1980}, 44, 1323.

\bibitem{bib16} \textsc{G. Rempe}, \textsc{H. Walther}, \textsc{ N. Klein},  Phys. Rev. Lett. \textbf{1987}, 58, 353.

\bibitem{bib17} \textsc{H.\,P. Breuer}, \textsc{F. Petruccione}, The Theory of Open Quantum Systems, Oxford University Press, Oxford \textbf{2002}.

\bibitem{bib18} \textsc{A.\,M. Kaufman}, \textsc{M.\,E. Tai}, \textsc{A. Lukin}, \textsc{M. Rispoli}, \textsc{R. Schittko}, \textsc{P.\,M. Preiss}, \textsc{M. Greiner},  Science \textbf{2016}, 353, 794.

\bibitem{bib19} \textsc{M.\,C. Arnesen},  \textsc{S. Bose}, \textsc{V. Vedral},  Phys. Rev. Lett. \textbf{2001}, 87, 017901.

\bibitem{bib20} \textsc{X.\,G. Wang},  Phys. Rev. A \textbf{2001}, 64, 012313.

\bibitem{bib21} \textsc{J.\,P. Reithmaier},   \textsc{G.\,Sek, A. L\"{o}ffler},  \textsc{C. Hofmann},  \textsc{S. Kuhn},  \textsc{S. Reitzenstein},  \textsc{L.\,V. Keldysh},  \textsc{V.\,D. Kulakovskii},  \textsc{T.\,L. Reinecke}, \textsc{A. Forchel}, Nature \textbf{2004}, 432, 197.

\bibitem{bib22} \textsc{E. Peter},  \textsc{P. Senellart},  \textsc{D. Martrou},  \textsc{A. Lema\^{i}tre}, \textsc{J. Hours},  \textsc{J.\,M. G\'{e}rard},  \textsc{J. Bloch}, Phys. Rev. Lett. \textbf{2005}, 95, 067401.

\bibitem{bib23} \textsc{H. Wichterich}, \textsc{M. J.Henrich}, \textsc{H.-P. Breuer},  \textsc{J. Gemmer},  \textsc{M. Michel}, Phys. Rev. E \textbf{2007}, 76, 031115.

\bibitem{bib24} \textsc{{\'A}. Rivas}, \textsc{A. D. K. Plato}, \textsc{S. F. Huelga}, \textsc{M. B. Plenio}, New. J. Phys. \textbf{2010}, 12, 113032.

\bibitem{bib25} \textsc{A. Levy}, \textsc{R. Kosloff}, EPL. \textbf{2014}, 107, 20004.

\bibitem{bib26} \textsc{J. O. Gonz{\'a}lez}, \textsc{L. A. Correa}, \textsc{G. Nocerino}, \textsc{J. P. Palao}, \textsc{D. Alonso}, \textsc{G. Adesso},  World Scientific Pub. Co Inc. \textbf{2017}, 4, 1740010.

\bibitem{bib27} \textsc{M. T. Mitchison}, \textsc{M. B. Plenio}, New J. Phys. \textbf{2018}, 20, 033005.

\bibitem{bib28} \textsc{G. D. Chiara}, \textsc{G. Landi}, \textsc{A. Hewgill}, \textsc{B. Reid}, \textsc{A. Ferraro}, \textsc{A. J. Roncaglia}, \textsc{M. Antezza}, New J. Phys. \textbf{2018}, 20, 113024.

\bibitem{bib29} \textsc{L. A. Correa}, \textsc{B. q. Xu}, \textsc{B. Morris}, \textsc{G. Adesso}, J. Chem. Phys. \textbf{2019}, 151, 094107.

\bibitem{bib30} \textsc{H.\,J. Carmichael}, \textsc{D.\,F. Walls}, J. Phys. A: Math. Nucl. Gen. \textbf{1973}, 6, 1552.

\bibitem{bib31} \textsc{M. Scala}, \textsc{B. Militello}, \textsc{A. Messina}, \textsc{J. Piilo}, \textsc{S. Maniscalco},  Phys. Rev. A \textbf{2007}, 75, 013811.

\bibitem{bib32} \textsc{Z. Wang}, \textsc{D.\,L. Zhou}, Phys. Rev. A \textbf{2014}, 89, 013809.

\bibitem{bib33} \textsc{ J.-Q. Liao}, \textsc{J.-F. Huang}, \textsc{L.-M. Kuang}, Phys. Rev. A \textbf{2011}, 83, 052110.

\bibitem{bib34} \textsc{L.\,A. Wu}, \textsc{D. Segal}, Phys. Rev. A \textbf{2011}, 84, 012319.

\bibitem{bib35} \textsc{S.-W. Li}, \textsc{C.\,Y. Cai}, \textsc{C.\,P. Sun}, Ann. Phys. \textbf{2015}, 360, 19.

\bibitem{bib36} \textsc{J. Eisert}, \textsc{M. Friesdorf}, \textsc{C. Gogolin}, Nat. Phys. \textbf{2015}, 11, 124.

\bibitem{bib37} \textsc{ J.-F. Huang}, \textsc{C.\,K. Law}, Phys. Rev. A \textbf{2015}, 91, 023806.

\bibitem{bib38} \textsc{C. Wang}, \textsc{X.-M. Chen}, \textsc{K.-W. Sun}, and \textsc{J. Ren}, Phys. Rev. A \textbf{2018}, 97, 052112.

\bibitem{bib39} \textsc{B. q. Guo}, \textsc{T. Liu}, \textsc{C. s. Yu}, Phys. Rev. E \textbf{2019}, 99, 032112.

\bibitem{bib40} \textsc{D. Braun}, Phys. Rev. Lett. \textbf{2002}, 89, 277901.

\bibitem{bib41} \textsc{M. Ikram}, \textsc{F.\,L. Li}, \textsc{M.\,S. Zubairy}, Phys. Rev. A \textbf{2007}, 75, 062336.

\bibitem{bib42} \textsc{M. Scala}, \textsc{R. Migliore}, \textsc{A. Messina},  J. Phys. A: Math. Theor. \textbf{2008}, 41, 435304.

\bibitem{bib43} \textsc{J. Li}, \textsc{G.\,S. Paraoanu}, New J. Phys. \textbf{2009}, 11, 113020.

\bibitem{bib44} \textsc{F. Benatti}, \textsc{R. Floreanini}, \textsc{U. Marzolino}, Phys. Rev. A \textbf{2010}, 81, 012105.

\bibitem{bib45} \textsc{L. Duan}, \textsc{H. Wang}, \textsc{Q.-H. Chen}, \textsc{Y. Zhao}, J. Chem. Phys. \textbf{2013}, 139, 044115.

\bibitem{bib46} \textsc{B. Bellomo}, \textsc{M. Antezza},  New J. Phys. \textbf{2013}, 15, 113052.

\bibitem{bib47} \textsc{G. Guarnieri}, \textsc{M. Kol\'{a}\v{r}}, \textsc{R. Filip}, Phys. Rev. Lett. \textbf{2018}, 121, 070401.

\bibitem{bib48} \textsc{ F. Tacchino}, \textsc{A. Auff\`{e}ves}, \textsc{M. F. Santos}, \textsc{D. Gerace}, Phys. Rev. Lett. \textbf{2018}, 120, 063604.

\bibitem{bib49} \textsc{Z. Wang}, \textsc{W. Wu}, \textsc{J. Wang}, Phys. Rev. A \textbf{2019}, 99, 042320.

\bibitem{bib50} \textsc{C.-H. Chou}, \textsc{T. Yu}, \textsc{B.\,L. Hu}, Phys. Rev. E \textbf{2008}, 77, 011112.

\bibitem{bib51} \textsc{L.\,S. Bishop}, \textsc{J.\,M. Chow}, \textsc{J. Koch}, \textsc{A.\,A. Houck}, \textsc{M.\,H. Devoret}, \textsc{E. Thuneberg}, \textsc{S.\,M. Girvin}, \textsc{R.\,J. Schoelkopf}, Nat. Phys. \textbf{2009}, 5, 105.

\bibitem{bib52} \textsc{J.\,P. Eckmann}, \textsc{C.\,A. Pillet}, \textsc{L. Rey-Bellet}, Commun. Math. Phys. \textbf{1999}, 201, 657.

\bibitem{bib53} \textsc{S. Trimper}, Phys. Rev. E \textbf{2006}, 74,051121.

\bibitem{bib54} \textsc{X.\,L. Huang}, \textsc{J.\,L. Guo}, \textsc{X.\,X. Yi}, Phys. Rev. A \textbf{2009}, 80, 054301.

\bibitem{bib55} \textsc{R.\,S. Johal}, Phys. Rev. E \textbf{2009}, 80, 041119.

\bibitem{bib55p5} \textsc{M. A. Nielsen}, \textsc{I. I. Chuang}, Quantum Computation and Quantum Information, United Kingdom at the University  Press, Cambridge \textbf{2000}.

\bibitem{bib56} \textsc{J. D. Cresser}, J. Mod. Opt.  \textbf{1992}, 39, 2187.

\bibitem{bib57} \textsc{G. Vidal}, \textsc{R. F. Werner}, Phys. Rev. A \textbf{2002}, 65, 032314.

\bibitem{bib58} \textsc{M.\,B. Plenio},  Phys. Rev. Lett. \textbf{2005}, 95, 090503.

\bibitem{bib60} \textsc{A. F. Kockum}, \textsc{A. Miranowicz}, \textsc{S. D. Liberato}, \textsc{S. Savasta}, \textsc{F. Nori}, Nat. Rev. Phys. \textbf{2019}, 1, 19.

\bibitem{bib61} \textsc{P. Forn-D\'{\i}az},  \textsc{ L. Lamata}, \textsc{E. Rico}, \textsc{J. Kono}, \textsc{E. Solano}, Rev. Mod. Phys. \textbf{2019}, 91, 025005.

\bibitem{bib64} \textsc{A. Smirne},  \textsc{ H.-P. Breuer}, \textsc{J. Piilo}, \textsc{B. Vacchini},  Phys. Rev. A  \textbf{2010}, 82, 062114.
\end{thebibliography}

\end{document}